\documentclass[pra, twocolumn,  superscriptaddress]{revtex4}
\usepackage{eurosym}
\usepackage{amssymb}
\usepackage{amsmath}
\usepackage{graphicx}
\usepackage{dcolumn}
\usepackage{color}
\usepackage{bm}
\usepackage{subfigure}
\usepackage{amsfonts}

\begin{document}

\preprint{APS/123-QED}
\title{ Detecting  the event of a single photon loss on quantum signals}
\author{ A. Mandilara}
\affiliation{Department of Physics, School of Science and Humanities, Nazarbayev
University, 53 Kabanbay Batyr Avenue, Nur-Sultan 010000, Kazakhstan}
\author{Y. Balkybek }
\affiliation{Department of Physics, School of Science and Humanities, Nazarbayev
University, 53 Kabanbay Batyr Avenue, Nur-Sultan 010000, Kazakhstan}
\author{V. M. Akulin}
\affiliation{Laboratoire Aim\'{e}-Cotton CNRS FRE 2038, L'Universit\'{e} Paris-Saclay et
L'\'{E}cole Normale Sup\'{e}rieure de Cachan, B\^{a}t. 505, Campus d'Orsay,
91405 Orsay Cedex, France}
\affiliation{Institute for Information Transmission Problems of the Russian Academy of
Science, 19 Bolshoy Karetny Per, Moscow, 127994, Russia.}

\begin{abstract}
We design a scheme for detecting a single photon loss from multi-modal quantum signals transmitted via a fiber or in free space.
 This consists of a
special type of  unitary coding transformation, the controlled-squeezing,
applied prior to the transmission on the signal composed by information and ancilla modes.  At the
receiver, the inverse unitary transformation is applied -decoding, and  the ancilla modes are measured via photon
detection. The outcome reveals whether a photon loss has occurred.
Distortion of the information part of the signal caused by an ancilla photon
loss can be corrected if the encoding transformation is appropriately selected. Loss of a photon from
the  information  part of the signal can be detected with the probability
exponentially close to unity. In contrast to the schemes of decoherence free subspaces and quantum error correction protocols, this methods allows one to make use of entire Hilbert  space dimensionality. We discuss possible ways of synthesizing the
required encoding-decoding transformations.
\end{abstract}

\maketitle

\section{Introduction \label{I}}

Photonic quantum information processing \cite{rev1, rev2} is a fast
advancing domain with break-through promises in the area of communication
security and computation. The main hindrance in the implementation of
photonic schemes is the irreversible process of photon losses which may
occur during propagation of a signal in free-space, fibers or integrated
photonic circuits. As a consequence many works have been dedicated to this subject suggesting various  elegant  solutions.
One  class of schemes has been deployed   aiming the realization of   noiseless amplification of quantum signals \cite{NA1, NA2, NA3, NA4, NA5, NA6}
in analogy to classical amplifiers. The most celebrated protection scheme against photon losses  
for  distributing  entangled pairs of photons  over long distances  is the one of quantum repeaters \cite{QR1, QR2, QR3}.
Finally, one main class of protection schemes  extends the ideas of quantum error-correcting codes for qubits \cite{Lidar} to photonic modes \cite{Kitaev, Grassl, Chuang, Munro, Banaszek, Mirrahimi, Girvin, Loock, Devoret, Shapiro}.

 In this work we propose an error-detection  scheme for photonic signals and we
focus on the single-photon losses that usually occur with a much larger
probability compared the higher order events \cite{Grassl, Chuang, Akulin}.
The task of error-detection is simpler  than the one of error-correction
and as consequence   the restrictions on the encoding subspace of the signal, present in every quantum error-correcting algorithm, can be ultimately lifted. Concerning the resources required for the scheme's implementation. These are of similar degree
of strain as for error-correcting codes \cite{Albert}, and we assume extra resources as $\chi^{\left(2\right)}$ non-linearity
or a controllable quantum ensemble of atoms that acts as a mediator to the fields.

\vspace{1cm}

 Detection of accidental irreversible changes is
well developed for classical signals \cite{Gala} being one of the main problems of
the coding theory. The task can be accomplished by adding to the main
signal an amount of complimentary information, so-called checksums, that
allows one to primarily detect the eventual distortion of the
transmitted information.  In this work  we extend the idea  to 
quantum optical signals and we devise a scheme that  
 detects uncontrolled errors of single photon-loss type on the information signal, by performing measurements on a complimentary ancilla system.
This task is implementable  by entangling the signal and ancilla at the encoding step and applying the inverse operation on the decoding step. 
However, due to this underlying entanglement  errors on the ancilla can generate errors on the quantum state of the information signal, so
 we carefully  adjust the scheme so that such a effect is correctable by simple unitary actions.
 On the other hand, if the information part of the quantum signal is 
subjected to an uncontrolled action, such an non-unitary error can only be
detected by our scheme while the distorted quantum information cannot be restored.

In our suggested  error detection scheme   quantum
information is encoded in  quantum states of multi-mode electromagnetic
fields. The input quantum optical signal $\left\vert \Psi \right\rangle
=\left\vert \Psi \right\rangle _{\mathcal{I}}\otimes \left\vert \Psi \right\rangle
_{\mathcal{A}} $ is composed of $K$ distinguishable information modes 
\begin{eqnarray}
\left\vert \Psi \right\rangle _{\mathcal{I}} &=&\sum_{n_{1}=0}^{N}\ldots
\sum_{n_{i}=0}^{N}\ldots \sum_{n_{K}=0}^{N}\psi _{n_{1},\ldots ,n_{i}\ldots
,n_{K}}  \notag \\
&&\times \left\vert n_{1}\right\rangle \ldots \left\vert n_{i}\right\rangle
\ldots \left\vert n_{K}\right\rangle  \label{first}
\end{eqnarray}%
with $\widehat{n}_{i}=\widehat{a}_{i}^{\dagger }\widehat{a}_{i}$ the $i$-th
field mode number operator for the $i$-th mode, and $M$ ancilla modes 
\begin{eqnarray}
\left\vert \Psi \right\rangle _{\mathcal{A}} &=&\sum_{m_{1}=0}^{N}\ldots
\sum_{m_{i}=0}^{N}\ldots \sum_{m_{K}=0}^{N}\varphi _{m_{1},\ldots
,m_{i}\ldots ,m_{M}} \notag  \\
&&\times \left\vert m_{1}\right\rangle \ldots \left\vert m_{i}\right\rangle
\ldots \left\vert m_{M}\right\rangle   \label{firstbis}
\end{eqnarray}%
with $\widehat{m}_{i}=\widehat{b}_{i}^{\dagger }\widehat{b}_{i}$. The $K+M$ modes in Eqs.(\ref{first})-(\ref{firstbis}) may stand for transverse spatial, angular momentum or polarization
modes.  The ancilla is initially set to the vacuum state  $
\left\vert 0\right\rangle _{\mathcal{A}}$ and the 
common quantum state $\left\vert \Psi \right\rangle$ is subject to the coding and decoding
transformations prior and after the propagation, respectively.
The key idea of the method lies on these coding/decoding actions which are performed with the help of either an energy
controlled-squeezing  operation, or a photon number parity controlled-squeezing
operation, -  unitary operations producing strong squeezing of the vacuum
states of the ancilla modes in function of either energy or the photon
number parity of the information signal modes. The first encoding scheme can
be synthesized in a relatively simple way, although has an important
shortage of restricting considerably the dimensionality of the Hilbert space
suitable for the encoding of the information. The second one can make use of
the entire dimensionality, but requires more complicated means for its implementation.

\begin{figure}[h]
\includegraphics
[width=0.4\textwidth]
{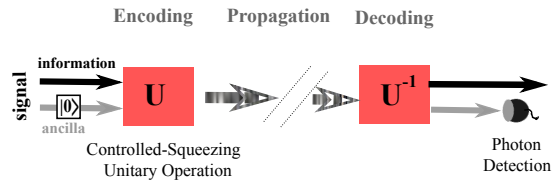} \vspace{1cm}
\caption{ A schematic representation of the four phases of the detection
scheme. \textit{a)} Encoding where the signal's information modes become entangled with
the ancilla modes. \textit{b)} Propagation of the `dressed' signal via the
medium. At this phase the incident of a photon loss may occur. \textit{c)}
Decoding. \textit{\ d)} Photon counting on each of the ancilla modes.}
\label{fig1}
\end{figure}

The paper is organized as follows. In Section~\ref{II} we describe a scheme
that is able to distinguish between two error events: no photon losses and a
single photon loss on the signal or ancilla. As shown in Fig.\ref{fig1}, it
includes the phases of encoding, propagation, decoding and detection. During
the encoding phase, the ancilla modes are getting entangled with the signal
via a unitary operation that we call \textit{energy controlled-squeezing}
operation. In Section~\ref{V} we consider the unitary operation that we call 
\textit{photon number parity controlled-squeezing} operation, which allows one to
encode the information in the entire Hilbert space of the photon modes, and
not only in the states corresponding a fixed number of the information
photons, as it is the case for the approach of Section~\ref{II}. In Section~\ref{III} 
we discuss  ways for synthesizing the encoding/decoding gates using 
standard methods of gates' decomposition and quantum control techniques. 

\section{The detection scheme \label{II}}

The general description of the amplitude damping in quantum channels \cite{Grassl, Chuang} is based on the theory of open quantum systems and 
employes the density matrix and Kraus operators formalism. However this
description drastically simplifies \cite{Banaszek} under the assumption of a
single photon loss and pure states input, where one can remain within the Hilbert space formalism
describing a photon loss in mode $i$ by simple action of the anihilaion
operator $\widehat{a}_{i}$ on the signal state $\left\vert \Psi
\right\rangle $, or even a photon loss from a superposition of modes given
by the collective photon loss operator 
\begin{equation}
\widehat{A}=\sum_{i=1}^{K}\alpha _{i}\widehat{a}_{i}  \label{uni}
\end{equation}%
with $\sum_{i=1}^{K}\left\vert \alpha _{i}\right\vert ^{2}=1$.

\vspace{1cm} 

\subsection{Effect of coding, photon losses and decoding on the quantum
field signal state}

Let us start with the unknown $K$-mode quantum state $\left\vert \Psi
\right\rangle _{\mathcal{I}}$ of the information system Eq.(\ref{first}) to be
transported to a final destination either through a fiber or via the free
space. To protect the signal from the uncontrolled photon loss of Eq.(\ref%
{uni}), we add a $M$-mode ancilla Eq.(\ref{firstbis}) in the vacuum state $
\left\vert 0\right\rangle _{\mathcal{A}}$ and encode this combined quantum system of $%
\left( K+M\right) $ modes by applying an entangling unitary transformation $%
\widehat{U}=e^{-i\widehat{S}\left( \widehat{a}_{i}^{\dagger },\widehat{a}%
_{i},\widehat{b}_{j}^{\dagger },\widehat{b}_{j}\right) }$, which we first
take in the form 
\begin{equation}
\widehat{U}_{ECS}=e^{-i\sum_{i,j=1}^{K,M}\gamma _{i,j}\left( \widehat{a}%
_{i}^{\dagger }\widehat{a}_{i}\right) \otimes \left( \widehat{b}%
_{j}^{\dagger }\widehat{b}_{j}^{\dagger }+\widehat{b}_{j}\widehat{b}%
_{j}\right) }.  \label{enco}
\end{equation}%
This operation produces squeezing of the ancilla modes controlled by linear combinations
of the photon number operators of the information modes and therefore we name it an \textsl{ energy controlled-squeezing} operation. Obviously the additional 
ancilla modes increase the number of possible photon loss channels since,
apart from the losses given by the operator $\widehat{A}$, one should now
take into account the ancilla losses given by the operator $\widehat{B}=\sum_{j=1}^{M}%
\beta _{j}\widehat{b}_{j}$ with $\sum_{j=1}^{M}\left\vert \beta
_{j}\right\vert ^{2}=1$. After the propagation, at the final destination,
the entangled signal is decoded by applying $\widehat{U}^{-1}=e^{i\widehat{S}%
\left( \widehat{a}_{i}^{\dagger },\widehat{a}_{i},\widehat{b}_{j}^{\dagger },%
\widehat{b}_{j}\right) }$, the inverse of the unitary operator (\ref{enco}),
and the state of the ancilla is subject to a measurement by detecting the
photon numbers in each of its $M$ modes.

Consider now three possible outcomes of the measurement process:\textit{ (i)} In the
case of no losses, the combined signal is restored intact after the
decoding, as a direct product $\left\vert \Psi \right\rangle =\left\vert
\Psi \right\rangle _{\mathcal{I}}\otimes \left\vert 0\right\rangle _{\mathcal{A}}$. As
consequence no photons are detected on any of the $M$ ancilla modes and the
signal state is not affected by these measurements. \textit{(ii)} In the case where a
photon has been lost from an ancilla mode, the state $\left\vert \Psi
\right\rangle =\widehat{U}_{ECS}^{-1}\widehat{B}\widehat{U}_{ECS}\left\vert \Psi
\right\rangle _{\mathcal{I}}\otimes \left\vert 0\right\rangle _{\mathcal{A}}$ reads 
\begin{equation}
\left\vert \Psi \right\rangle =\Delta \sum_{j=1}^{M}\beta _{j}\sinh \left(
2\sum_{i=1}^{K}\gamma _{i,j}\left( \widehat{a}_{i}^{\dagger }\widehat{a}
_{i}\right) \right) \left\vert \Psi \right\rangle _{\mathcal{I}}\otimes \widehat{b}
_{j}^{\dagger }\left\vert 0\right\rangle _{\mathcal{A}}  \label{vi}
\end{equation}
with one of the ancilla modes state being in the first
excited state. Here $\Delta $ is a normalization factor associated with the
non-unitary action of the operators $\widehat{b}
_{j}$. The measurement of a photon at the $j$
ancilla mode  projects the signal state to 
\begin{equation}
\left\vert \Psi ^{\prime }\right\rangle _{\mathcal{I}}=\Delta' \sinh \left(
2\sum_{i=1}^{K}\gamma _{i,j}\left( \widehat{a}_{i}^{\dagger }\widehat{a}
_{i}\right) \right) \left\vert \Psi \right\rangle _{\mathcal{I}}~,  \label{vi2}
\end{equation}
with $\Delta'$ a normalization factor associated with the measurement induced state re-
duction.
Eq.(\ref{vi2}) implies that the subspace of the entire Hilbert space of the
information signal corresponding to the fixed linear combination of the
photon numbers $\sum_{i=1}^{K}\gamma _{i,j}n_{i}=\mathrm{const}$ remains
intact.\textit{ (iii)} In the case where a photon has been lost from the information
system one finds the final state $\left\vert \Psi \right\rangle =\widehat{U}_{ECS}
^{-1}\widehat{A}\widehat{U}_{ECS}\left\vert \Psi \right\rangle _{\mathcal{I}}\otimes
\left\vert 0\right\rangle _{\mathcal{A}}$ as 
\begin{equation}
\left\vert \Psi \right\rangle =\sum_{i=1}^{K}\alpha _{i}\widehat{a}
_{i}\left\vert \Psi \right\rangle _{\mathcal{I}}\otimes e^{-i\sum_{j=1}^{M}\gamma
_{i,j}\left( \widehat{b}_{j}^{\dagger }\widehat{b}_{j}^{\dagger }+\widehat{b}
_{j}\widehat{b}_{j}\right) }\left\vert 0\right\rangle _{\mathcal{A}},  \label{vi3}
\end{equation}
where the ancilla modes are conditionally squeezed on the loss of a photon.

For large squeezing parameters $\gamma _{i,j}$, in the case \textit{(iii)}, by
measurement on the number of photons in the ancilla modes, apart from the
event of no photon registered having a vanishing probability 
\begin{eqnarray}
P_{0} &=&\sum_{i=1}^{K}\left\vert \alpha _{i}\right\vert ^{2}\left.\right._{\mathcal{A}}\left\vert
\left\langle 0\right\vert e^{-i\sum\limits_{j=1}^{M}\gamma _{i,j}\left( 
\widehat{b}_{j}^{\dagger }\widehat{b}_{j}^{\dagger }+\widehat{b}_{j}\widehat{
b}_{j}\right) }\left\vert 0\right\rangle _{\mathcal{A}}\right\vert ^{2}  \label{Po} \\
&=&\sum_{i=1}^{K}\left\vert \alpha _{i}\right\vert ^{2}\prod\limits_{j=1}^{M}
\frac{2e^{-\left\vert \gamma _{i,j}\right\vert }}{1+e^{-\left\vert \gamma
_{i,j}\right\vert }}\sim e^{-M\overline{\left\vert \gamma _{i,j}\right\vert }
},  \notag
\end{eqnarray}
one detects a large even typical number $m_{j}\sim \sum_{i=1}^{K}\alpha
_{i}\exp 2\left\vert \gamma _{i,j}\right\vert $ of  photons \cite{Schleich}
while the joint photon count distribution \cite{Manko} for the ancilla modes reads 
\begin{eqnarray}
P_{\left\{ m\right\} } &=&\sum_{i=1}^{K}\left\vert \alpha _{i}\right\vert
^{2}\prod\limits_{j=1}^{M}\left\vert \left\langle m_{j}\right\vert
e^{-i\gamma _{i,j}\left( \widehat{b}_{j}^{\dagger }\widehat{b}_{j}^{\dagger
}+\widehat{b}_{j}\widehat{b}_{j}\right) }\left\vert 0\right\rangle
\right\vert ^{2}  \label{joint} \\
&=&\sum_{i=1}^{K}\left\vert \alpha _{i}\right\vert ^{2}\prod\limits_{j=1}^{M}%
\frac{m_{j}!\sqrt{1-\tanh ^{2}\frac{\gamma _{i,j}}{2}}\tanh ^{m_{j}}\frac{%
\gamma _{i,j}}{2}}{\left( m_{j}/2\right) !\left( m_{j}/2\right) !2^{m_{j}}} 
\notag
\end{eqnarray}%
with all $m_{j}$ even numbers.

\subsection{Structuring the coupling elements in the energy controlled-squeezing Hamiltonian}

\begin{figure}[h]
\includegraphics
[width=0.3\textwidth]
{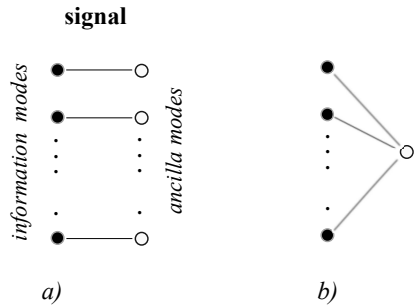} \vspace{1cm}
\caption{ The principal coupling schemes investigated in this work. \textit{a)} The
number of information modes equals the number of ancilla modes  and the coupling is
one-to-one as described in Eq.(\protect\ref{g1}). \textit{b)} There is only
one ancilla mode coupled with each one of the signal's information modes in a symmetric way,
Eq.(\protect\ref{g2}).}
\label{fig2}
\end{figure}

Thus far, no assumptions have been imposed on the matrix $\gamma _{i,j}$ in Eq.(\ref{enco}),
other than big amplitude resulting to strong squeezing of the ancilla mode states. Here, we structure
the elements of the coupling matrix imposing two different types of symmetric coupling
presented in Fig.~\ref{fig2}. Each one serves a different aim: we want to either
\textit{(i)} make the scheme working in a way resembling the classical error
syndromes, or \textit{(ii)} maximize the dimension of the error-tolerant subspace in
the Hilbert space of the signal information quantum state thus augmenting the quantum
channel capacity.

\vspace{1cm} 

Let's first explore the case \textit{(i)} where one just wants to identify the information
mode $i$ where a photon has been lost. To this end, we set $M=K$ and require
that each of the signal modes is entangled with just one ancilla mode via a
matrix 
\begin{equation}
\gamma _{i,j}=\gamma \delta _{i,j}~,  \label{g1}
\end{equation}%
with $\gamma $ large positive number, Fig.~\ref{fig2}~(a). In view of the Eq.(
\ref{vi3}) the detection of two or more  photons on the ancilla mode $j$ will
indeed unambiguously indicate that a photon has been lost from the signal
mode $j$.

Still, the most logical use of the detection scheme for heralded quantum
communication purposes \textit{(ii)} requires a different structuring of the matrix $
\gamma _{i,j}$. In fact, according to Eq.(\ref{vi2}), the coupling Eq.(\ref%
{g1}) will distort the signal $\left\vert \Psi \right\rangle $ in the case
of an ancilla photon loss, thus leaving no error tolerant subspaces in the
Hilbert space of information signal. So the choice of the matrix as in Eq.(%
\ref{g1}) is not convenient for transmission of quantum information, since
the detection scheme results in a high probability of  signal distortion. 
This observation takes us to a second proposal for the coupling matrix as 
\begin{equation}
\gamma _{i,j}=\gamma k_{i,j}~~\mathrm{with}~k_{i,j}=k_{i^{\prime
},j},~\forall i,i^{\prime }.  \label{g2}
\end{equation}%
In the simplest case shown in Fig.~\ref{fig2}~(b), one can assume $M=1$. With
such a coupling one can no longer identify which one of the modes $i$ has
experienced loss of photon, but for a state $\left\vert \Psi \right\rangle $
comprising only the states with an identical total number of photons, the
signal will not be affected by the ancilla photon loss, in compliance with
Eq.(\ref{vi2}). In the simplest version of such a scenario, the subspace of
one photon can be employed for the encoding of the information as: 
\begin{eqnarray}
\left\vert \mathbf{0}\right\rangle  &=&\left\vert 1\right\rangle \left\vert
0\right\rangle \dots \left\vert 0\right\rangle   \notag \\
\left\vert \mathbf{1}\right\rangle  &=&\left\vert 0\right\rangle \left\vert
1\right\rangle \dots \left\vert 0\right\rangle   \notag \\
&&\ldots   \notag \\
\left\vert \mathbf{K}\right\rangle  &=&\left\vert 0\right\rangle \left\vert
0\right\rangle \dots \left\vert 1\right\rangle .  \label{code}
\end{eqnarray}
Such an encoding on product states is more accessible   than the coding over entangled states in the habitual error correcting codes, but still this is quite restrictive.

\section{Coding with a Parity controlled-squeezing operation \label{V}}

The coding based on the energy controlled-squeezing Eq.(\ref{enco})
obviously reduces the effective dimension of the encoding space from the
full dimension $N^{K}$ to just $K$ at most (see Eq.(\ref{code})), and hence a natural question
arises: can one  considerably augment this dimension by another choice of the
action $\widehat{S}\left( \widehat{a}_{i}^{\dagger },\widehat{a}_{i},
\widehat{b}_{j}^{\dagger },\widehat{b}_{j}\right) $ entering the coding
transformation $\widehat{U}=e^{-i\widehat{S}\left( \widehat{a}_{i}^{\dagger
},\widehat{a}_{i},\widehat{b}_{j}^{\dagger },\widehat{b}_{j}\right) }$? The
answer is positive: the coding action 
\begin{equation}
\widehat{S}=\Gamma \sum_{j=1}^{M}\widehat{\Pi }_{j}\otimes \left( \widehat{b}%
_{j}^{\dagger }\widehat{b}_{j}^{\dagger }+\widehat{b}_{j}\widehat{b}%
_{j}\right)   \label{encoP}
\end{equation}%
allows one to employ full dimensionality $N^{K}$ for the information
transmission. Here $\widehat{\Pi }_{j}=\left( -1\right)
^{\sum_{i=1}^{K}\gamma _{i,j}\widehat{a}_{i}^{\dagger }\widehat{a}_{i}}$
with $\gamma _{i,j}=0,1$ is the photon number parity operator for the
modes of the information system coupled to the ancilla
mode $j$.  We denote the unitary operation
corresponding to the action Eq.(\ref{encoP}) by $\widehat{U}_{PCS}$, and henceforth call it a \textit{parity controlled-squeezing} operation. 

We proceed by investigating  the encoding/decoding effect of $\widehat{U}_{PCS}$ taking the coupling matrix as in Eq.(\ref{g1}), Fig.~\ref{fig2}~(a).

\subsection{The outcome in the case of a photon loss on the ancilla modes}

 Employing $\widehat{U}_{PCS}$ as encoding operation,  a photon loss from the $j$ ancilla mode results in the signal
state 
\begin{equation}
\left\vert \Psi ^{\prime }\right\rangle_{\mathcal{I}} =\Delta ^{\prime
}\sinh \left( 2\Gamma \widehat{\Pi }_{j}\right) \left\vert
\Psi \right\rangle_{\mathcal{I}} ~,  \label{vi5}
\end{equation}
by  analogy to Eq.(\ref{vi2}).
By noting that $\sinh \left( x\right) $ is an odd function, one immediately
finds that $\left\vert \Psi ^{\prime }\right\rangle_{\mathcal{I}} =\widehat{\Pi }%
_{j}\left\vert \Psi \right\rangle_{\mathcal{I}} $.
This implies that  the states with an even number of photons in the $j$
information mode  coupled to the $j$ ancilla mode which has lost a
photon remain intact, while those having an odd number of photons experience
a $\pi $ phase shift. This transformation of the information system is
reversible, one simply needs to know which one of the ancilla mods $j$ has
lost the photon and apply the unitary operation $\widehat{\Pi }_{j}=e^{i \pi \widehat{a}_{j}^{\dagger }\widehat{a}_{j}}$ in 
accordance. This piece of  information is readily available from the ancilla photon
measurement -- the mode that has lost a photon during the stage of
propagation gets an extra photon after the decoding transformation as in Eq.(\ref{vi})), and this
 photon can be detected.

\subsection{The outcome in the case of a photon loss on the information modes}

In the case of a photon loss from the state mode $i$, the analogue to Eq.(\ref{vi3}) is
\begin{equation}
\left\vert \Psi ^{\prime }\right\rangle =\sum_{i=1}^{K}\alpha _{i} 
\widehat{a}_{i}\left\vert \Psi \right\rangle_{\mathcal{I}} \otimes  e^{-2 i\Gamma
\widehat{\Pi }_{i}\otimes \left( \widehat{b}_{i}^{\dagger }
\widehat{b}_{i}^{\dagger }+\widehat{b}_{i}\widehat{b}_{i}\right)  }\left\vert 0\right\rangle_{\mathcal{A}}.  \label{statepait}
\end{equation}
This implies that the photon loss results in a strong squeezing on the corresponding
ancilla mode $i$, such that the distribution
of the even photon number states in this mode reads 
\begin{equation}
P_{\left\{ m\right\} }=\sum_{i=1}^{K}\left\vert \alpha _{i}\right\vert
^{2}\frac{m_{i}!\sqrt{
1-\tanh ^{2}\Gamma }\tanh ^{m_{i}}\Gamma }{\left( m_{i}/2\right) !\left(
m_{i}/2\right) !2^{m_{i}}}.  \label{jointparity}
\end{equation}
Deriving Eq.(\ref{jointparity}) from Eq.(\ref{statepait}) we have employed
 the well-known fact that the photon distribution $P\left( m\right)
=\left\vert \left\langle m\right\vert e^{i\Gamma \left( \widehat{b}^{\dagger
}\widehat{b}^{\dagger }+\widehat{b}\widehat{b}\right) }\left\vert
0\right\rangle \right\vert ^{2}$ in the squeezed vacuum state of a single
mode 
\begin{equation*}
P\left( m\right) =\left( 1+\left( -1\right) ^{m}\right) \frac{m!\sqrt{%
1-\tanh ^{2}\Gamma }\tanh ^{m}\Gamma }{\left( m/2\right) !\left( m/2\right)
!2^{m+1}}
\end{equation*}%
is independent of the sign of the squeezing parameter. For such a
distribution, typical numbers of photons  in the $i$ ancilla mode is large,
being of the order of $\left\langle m_{i}\right\rangle \sim $exp$\left(
2\Gamma \right)$.
This implies that once a photon has been lost from a mode $i$ of the
information system, on  the ancilla mode $i$ the vacuum state gets strongly squeezed and a large even
number of photons can be detected  as a consequence. 

One can easily extend the above results to the
case where the information mode $i$ is coupled to more than one ancilla modes $j$ via a coupling matrix
$\gamma_{i,j}=0,1$. In such case the Eqs.(\ref{statepait}), (\ref{jointparity}) generalize to 
\begin{equation}
\left\vert \Psi ^{\prime }\right\rangle =\sum_{i=1}^{K}\alpha _{i}e^{i\Gamma
\sum_{j=1}^{M}\widehat{\Pi }_{j}\otimes \left( \widehat{b}_{j}^{\dagger }%
\widehat{b}_{j}^{\dagger }+\widehat{b}_{j}\widehat{b}_{j}\right) \left(
1-\left( -1\right) ^{\gamma _{i,j}}\right) }\left\vert 0\right\rangle_{\mathcal{A}} 
\widehat{a}_{i}\left\vert \Psi \right\rangle_{\mathcal{I}} ,  \label{statepait2}
\end{equation}
and
\begin{equation}
P_{\left\{ m\right\} }=\sum_{i=1}^{K}\left\vert \alpha _{i}\right\vert
^{2}\prod\limits_{j/\left\{ \gamma _{i,j}=1\right\} }\frac{m_{j}!\sqrt{%
1-\tanh ^{2}\Gamma }\tanh ^{m_{j}}\Gamma }{\left( m_{j}/2\right) !\left(
m_{j}/2\right) !2^{m_{j}}}  \label{jointparity2}
\end{equation}
where the product is taken over all states $j$ for which $\gamma _{i,j}=1$. In such case the 
 the loss of a photon on an information mode, transmits the squeezing effect on more than one
ancillary modes and the error can  more efficiently be detected.

\section{How to produce the controlled-squeezing operations?\label{III}}

We now address the question on the kind of the physical interaction which can
result in the controlled-squeezing coding actions Eq.(\ref{enco}),(\ref{encoP}), and we propose
specific quantum control scenarios and decompositions that can guide their realization.

\subsection{The energy controlled-squeeze gate }

We start with the operation of Eq.(\ref{enco}). The relevant detection scheme is
based on the possibility of implementing the two-mode gate 
\begin{equation}
\widehat{G}_{ECS}=e^{-i\gamma \left( \widehat{a}^{\dagger }\widehat{a}\right)
\otimes \left( \widehat{b}^{\dagger }\widehat{b}^{\dagger }+\widehat{b}%
\widehat{b}\right) }  \label{enco0}
\end{equation}%
between a signal mode $\widehat{a}$ and an ancilla mode $\widehat{b}$. We
call this gate an \textit{energy controlled-squeezing} gate since with its
application the mode $\widehat{b}$ is getting squeezed with a parameter that
depends linearly on the number of photons of the mode $\widehat{a}$. 

The coupling Hamiltonian in Eq.(\ref{enco0}) is a sort of quartic Kerr
nonlinearity, which may either not be present in the optical setting, or be
rather small making the required coding transformation Eq.(\ref{enco0})
difficult to achieve in practice. Alternatively, one can think about the
quantum signal mode energy $\widehat{a}^{\dagger }\widehat{a}$ controlling
via a nonlinear interaction $\chi ^{(2)}$ a coherent pump $E\sim \chi ^{(2)}%
\widehat{a}^{\dagger }\widehat{a}\mathcal{E}$ of a parametric generator,
which excites the ancilla modes by a Hamiltonian $E\left( \widehat{b}%
^{\dagger }\widehat{b}^{\dagger }+\widehat{b}\widehat{b}\right) $, where $%
\mathcal{E}$ is a strong classical field. However, this approach requires
complete conversion of the coherent pumping field photons to the photons of
the squeezed vacuum state of the ancilla, which can be guaranteed only for
the case of virtual pumping photons thus again implying a weak coupling
between modes $\widehat{a}$ and $\widehat{b}$. Therefore, we conclude that the realization of
such a gate by  straightforward quantum optical methods is rather inefficient and in the following
we adopt another methodology,  relying on the recent progress that has been made towards the implementation of cubic phase gates \cite{Sanders, Weedbrook}.

We proceed by proving that  the gate in Eq.(\ref{enco0}), can  be implemented on the basis of
single-mode cubic phase gates  and Gaussian 
operations. For the purpose, it is more convenient to work with the
canonical ``position'' and ``momentum'' mode operators defined for the signal
mode as $\widehat{q}_{a}=\frac{\widehat{a}+\widehat{a}^{\dagger }}{\sqrt{2}}$%
and $\widehat{p}_{a}=-i\frac{\widehat{a}-\widehat{a}^{\dagger }}{\sqrt{2}}$,
respectively, and for the ancilla mode $\widehat{b}$ as $\widehat{q}_{b}=%
\frac{1}{\sqrt{2}}\left( \widehat{b}+\widehat{b}^{\dagger }\right) $ and $%
\widehat{p}_{b}=-i\frac{1}{\sqrt{2}}\left( \widehat{b}-\widehat{b}^{\dagger
}\right) $. We  have also invoked here the scaled units where
the oscillator frequency equals unity. The commutation relation now reads $%
\left[ \widehat{q}_{l},\widehat{p}_{m}\right] =i\delta _{lm}$, while Eq.(\ref%
{enco0}) can be re-written as 
\begin{equation}
\widehat{G}_{ECS}=e^{-i\frac{\gamma }{2}\left( \widehat{q}_{a}^{2}+\widehat{p}%
_{a}^{2}\right) \otimes \left( \widehat{q}_{b}^{2}-\widehat{p}%
_{b}^{2}\right) ~.}  \label{enco1}
\end{equation}

Let us start with the two-mode Gaussian operation: 
\begin{equation}
\widehat{U}_0=e^{i \left(\widehat{q}_a+\widehat{p}_a\right)\otimes\left(%
\widehat{q}_b+\widehat{p}_b\right).}  \label{ba}
\end{equation}
By applying the single-qubit phase gates $\widehat{U}=\exp {i \lambda_1 
\widehat{q}^3_a}$, $\widehat{U}=\exp {i \mu_1 \widehat{q}^3_b}$ and their
inverses on Eq.(\ref{ba}), 
\begin{equation}
\widehat{U}_1=e^{i\lambda_1 \left(\widehat{q}_a\right)^3\otimes \widehat{I}%
}e^{i \mu_1 \widehat{I}\otimes \left(\widehat{q}_b\right)^3}\widehat{U}_0
e^{-i\lambda_1 \left(\widehat{q}_a\right)^3\otimes \widehat{I}}e^{-i \mu_1 
\widehat{I}\otimes \left(\widehat{q}_b\right)^3},
\end{equation}
one arrives into a quartic generating Hamiltonian 
\begin{equation}
\widehat{U}_1=e^{i\left(\widehat{q}_a+\widehat{p}_a -3 \lambda_1 \left(%
\widehat{q}_a\right)^2 \right)\otimes\left(\widehat{q}_b+\widehat{p}_b-3
\mu_1 \left(\widehat{q}_b\right)^2 \right)}.  \label{u1}
\end{equation}
where we have used the identity $\widehat{U}^{-1} e^{i \widehat{H}} \widehat{%
U} =e^{i \widehat{U}^{-1}\widehat{H} \widehat{U}}$.

One can then easily reduce Eq.(\ref{u1}) to Eq. (\ref{enco1}) by a sequence
of single-mode Gaussian operations. The simplest scenario is the following
sequence of transformations 
\begin{equation}
\widehat{U}_{2}=\left( \widehat{U}_{2a}\otimes \widehat{U}_{2b}\right) 
\widehat{U}_{1}\left( \widehat{U}_{2a}^{\dagger }\otimes \widehat{U}%
_{2b}^{\dagger }\right)   \label{u2}
\end{equation}%
where 
\begin{eqnarray}
\widehat{U}_{2a} &=&e^{i\lambda _{6}\widehat{p}_{a}+i\lambda _{5}\widehat{p}%
_{a}^{2}}e^{-i\lambda _{4}\widehat{q}_{a}^{2}}e^{i\lambda _{3}\widehat{p}%
_{a}^{2}}e^{i\lambda _{2}\widehat{q}_{a}^{2}} \\
\widehat{U}_{2b} &=&e^{i\mu _{6}\widehat{p}_{b}+i\mu _{5}\widehat{p}%
_{b}^{2}}e^{i\mu _{4}\widehat{q}_{b}^{2}}e^{i\mu _{3}\widehat{p}%
_{b}^{2}}e^{i\mu _{2}\widehat{q}_{b}^{2}}.
\end{eqnarray}%
One then needs to solve a linear system of equations to identify the
parameters $\lambda $'s and $\mu $'s such that $\widehat{U}_{2}$, Eq.(\ref%
{u2}), matches the wanted operation $\widehat{G}_{ECS}$.

It is important to mention that the decomposition of the  energy controlled-squeezing
gate that we propose in this work is  indicative. It might be that there
is a shorter (approximate) decomposition using standard compilation methods
for bosonic gates \cite{vL, xanadu}.

\subsection{The photon parity controlled-squeezing  gate}

We now turn to the construction of the coding transformation 
Eq.(\ref{encoP}), considering the simplest case of just one information mode and  one
ancilla mode, such that the required coding transformation is given by
\begin{equation}
\widehat{G}_{PCS}=\exp \left[ -i\Gamma e^{i\pi \widehat{a}^{\dagger }\widehat{a}%
}\otimes \left( \widehat{b}^{\dagger }\widehat{b}^{\dagger }+\widehat{b}%
\widehat{b}\right) \right] ,  \label{enco2}
\end{equation}
which we name the parity controlled-squeezing gate.
We show how one can synthesize such a gate with the help of an
intermediate system with $SU(2)$ symmetry, such as a collection of non-interacting two-level systems which can be described in terms
of three Pauli operators $\widehat{\sigma }_{x},\widehat{\sigma }_{y},$ and $%
\widehat{\sigma }_{z}$  entering the polarization, dispersion, and the
population inversion components of the Bloch vector, respectively.

Let $\widehat{V}_{i,tl}=\mu \widehat{a}^{\dagger }\widehat{a}\widehat{\sigma 
}_{z}$ be the interaction Hamiltonian between the intermediate system and the
information system. Physically this corresponds to the Stark shift in the two-level systems
induced by the electric field of the information photons. Let also  the
 Hamiltonian $\widehat{V}_{a,tl}=\kappa \left( \widehat{b}%
^{\dagger }\widehat{b}^{\dagger }+\widehat{b}\widehat{b}\right) \widehat{%
\sigma }_{x}$ describe the interaction of the intermediate system with the ancilla, implying that the
total polarization of the two-level systems parametrically pumps the ancilla
photon mode. We assume that the coupling energy $\mu $ can be done both
positive and negative by a proper choice of the intermediate system
parameters. We also assume the intermediate system  initially in the eigenstate $\left\vert 1\right\rangle _{\mathcal{TL}}$ of the operator $\widehat{\sigma }_{x}$ with eigenvalue $1$, such that initially, the entire
system is in the quantum state $\left\vert \Psi \right\rangle _{\mathcal{I}} \otimes \left\vert
0\right\rangle _{\mathcal{A}}\otimes\left\vert 1\right\rangle _{\mathcal{TL}}$. 

In order to synthesize the required coding transformation, we first apply
the interaction $\widehat{V}_{i,tl}$ for a time interval $\pi /2\mu $, we
then apply the interaction $\widehat{V}_{a,tl}$ during a time interval $
\Gamma /\kappa $, and finally apply the interaction $-\widehat{V}_{i,tl}$
during a time interval $\pi /2\mu $, thus bringing the entire compound system
to the quantum state       
\begin{eqnarray}
\left\vert \Psi \right\rangle _{\mathcal{C}} = e^{i\frac{\pi}{2} \widehat{a
}^{\dagger }\widehat{a}\otimes\widehat{I}\otimes\widehat{\sigma }_{z}}e^{-i\Gamma \widehat{I}\otimes\left( \widehat{b}^{\dagger }\widehat{b}%
^{\dagger }+\widehat{b}\widehat{b}\right)\otimes\widehat{\sigma }_{x} }e^{-i\frac{\pi}{2} \widehat{a
}^{\dagger }\widehat{a}\otimes\widehat{I}\otimes\widehat{\sigma }_{z}} \notag \\
\times \left\vert \Psi \right\rangle
_{\mathcal{I}}\otimes\left\vert 0\right\rangle _{\mathcal{A}}\otimes\left\vert 1\right\rangle _{\mathcal{TL}} 
\notag \\
=e^{-i\Gamma \widehat{I}\otimes\left( \widehat{b}^{\dagger }\widehat{b}%
^{\dagger }+\widehat{b}\widehat{b}\right)\otimes \widehat{I}} e^{i\frac{\pi}{2} \widehat{a
}^{\dagger }\widehat{a}\otimes\widehat{I}\otimes\widehat{\sigma }_{z}} \left(\widehat{I}\otimes\widehat{\sigma }_{x}\otimes\widehat{I}\right) \notag \\
\times e^{-i\frac{\pi}{2} \widehat{a
}^{\dagger }\widehat{a}\otimes\widehat{I}\otimes\widehat{\sigma }_{z}}\left\vert \Psi \right\rangle _{\mathcal{I}}\otimes\left\vert
0\right\rangle _{\mathcal{A}}\otimes\left\vert 1\right\rangle _{TL}.  \label{enco3}
\end{eqnarray}
The operator $e^{i\frac{\pi}{2} \widehat{a}^{\dagger }\widehat{a}\otimes \widehat{\sigma }_{z}}%
\left(\widehat{I}\otimes\widehat{\sigma }_{x}\right)e^{-i\frac{\pi}{2} \widehat{a}^{\dagger }\widehat{a}\otimes \widehat{\sigma }_{z}}=\cos \left( \pi \widehat{a}^{\dagger }\widehat{a}\right)\otimes \widehat{\sigma }_{x} +i
\sin \left( \pi \widehat{a}^{\dagger }\widehat{a}
\right) \otimes\widehat{\sigma }_{y}$ with the operator $\sin \left( \pi \widehat{a}^{\dagger }\widehat{%
a}\right) $ vanishing for all number states of the information system and
their linear combinations thus yielding
\begin{equation}
\left\vert \Psi \right\rangle _{\mathcal{C}}=\left\vert 1\right\rangle _{\mathcal{TL}}\otimes
e^{-i\Gamma \cos \left( \pi \widehat{a}^{\dagger }\widehat{a}\right) \otimes\left( \widehat{b}^{\dagger }\widehat{b}^{\dagger }+\widehat{b}%
\widehat{b}\right) 
}\left\vert \Psi \right\rangle _{\mathcal{I}}\otimes\left\vert 0\right\rangle _{\mathcal{A}}.
\label{enco4}
\end{equation}
This means that the intermediate system is not anymore entangled with the
information system plus ancilla signal compound, while the latter experience the
required coding transformation. By setting the intermediate system's initial state to the
eigenstate $\left\vert 1\right\rangle _{\mathcal{TL}}$ of the operator $\widehat{%
\sigma }_{x}$ corresponding to the eigenvalue $-1$, one obtains the decoding
transformation%
\begin{equation}
\left\vert \Psi \right\rangle _{\mathcal{C}}=\left\vert -1\right\rangle
_{\mathcal{TL}}\otimes e^{i\Gamma \cos \left( \pi \widehat{a}^{\dagger }\widehat{a}\right) \otimes\left( \widehat{b}^{\dagger }\widehat{b}^{\dagger }+%
\widehat{b}\widehat{b}\right)}\left\vert \Psi \right\rangle _{\mathcal{I+A}},  \label{enco5}
\end{equation}
while the intermediate system remains disentangled from the rest.

\section{Discussion \label{IV}}

We have proposed a general scheme for detecting the event of a single photon loss on 
photonic signals composed by discrete  modes propagating via a
quantum channel. This comes as a workable alternative to well
established quantum error-correcting algorithms for bosonic modes,
imposing less or even, no restrictions on the encoding space of the signal and on the number of ancillary modes.
On the other hand, the suggestion of this current work is not compatible with quantum photonic computing processing
and it is only helpful for heralded quantum communication.

 The proposal is based on the use of ancilla modes which
`dress' the signal during the propagation in such a way that: \textit{a)} a
signal which did not loose any photons remains intact, \textit{b)} a
photon loss from the information system can be identified by photon detection
on the ancilla modes, and \textit{c)} a loss of a photon on the ancilla 
is also detectable but leaving intact only certain subspaces in the signal's Hilbert space. The
dimensionality of these subspaces depend on the form of the coding
transformation, reaching the full dimension for special settings.

The main tool of the approach is the conditional squeezing of the ancilla
modes, which in the regime of strong squeezing  and in the situation where  one of the information
carrying photons is lost, this results in a strong signal indication, i.e., the presence of an even number of photons
on the ancilla modes.  In contrast, the loss of an ancilla photon, after
decoding, yields the presence of a single photon on that mode. In the generic case, the
resulting state of the information system also gets distorted, but for
certain coding transformations this distortion is reversible and can be eliminated by proper
unitary correction.

One can synthesize the required encoding/decoding transformations, either  with
the help of  quantum control techniques or by deriving a decomposition over a series of known photonic gates. We have suggested a decomposition based
on  Gaussian operations and cubic phase gates, which is closely related with the experimentally
realizable settings. We have also proposed a way for the implementation of the
transformation which offers opportunities for complete preservation of the
signal, synthesized by a quantum control protocol.

\section*{Acknowledgment}

AM and YB acknowledge financial support by the Nazarbayev University ORAU
grant ``Dissecting the collective dynamics of arrays of superconducting
circuits and quantum metamaterials'' (no. SST2017031) and the MES RK
state-targeted program BR$05236454$.


\end{document}